\documentclass[pre,aps,showpacs,longbibliography,twocolumn]{revtex4-1}

\usepackage{rotating}		
\usepackage{amsmath}	
\usepackage{amssymb}	
\usepackage{color}
\usepackage[hidelinks]{hyperref}
\hypersetup{  
colorlinks=true,
citecolor=blue} 
\usepackage{bm}

\newcommand{\f}[2]{
		\mathchoice%
			{\dfrac{#1}{#2}}
	    	{\dfrac{#1}{#2}}
			{\frac{#1}{#2}}
			{\frac{#1}{#2}}
}

\newcommand{\dd}{\mathrm{d}}

\newcommand{\Sum}[2]{\sum\limits_{#1}^{#2}}

\newcommand{\moyenne}[1]{\left\langle #1\right\rangle}

\newcommand{\ddf}[3][]{%
        \ifthenelse{\equal{#1}{}}{%
                \ensuremath{\f{\dd#2}{\dd#3}}%
        }{%
                \ensuremath{\f{\dd^{#1}#2}{\dd{#3}^{#1}}}%
        }%
}

\newcommand{\Dp}[3][]{
        \ifthenelse{\equal{#1}{}}{%
                \ensuremath{\f{\partial#2}{\partial#3}}%
        }{%
                \ensuremath{\f{\partial^{#1}#2}{\partial{#3}^{#1}}}%
        }%
}

\newcommand{\ct}{{\mathrm{ct}}}

\begin{document}

\title{Efficiency fluctuations of small machines with unknown losses}

\author{Hadrien Vroylandt, Anthony Bonfils and Gatien Verley}
\affiliation{Laboratoire de Physique Th\'eorique (UMR8627), CNRS, Univ. Paris-Sud, Universit\'e Paris-Saclay, 91405 Orsay, France}
\date{\today}

\begin{abstract}

The efficiency statistics of a small thermodynamic machine has been recently investigated assuming that the total dissipation is a linear combination of two currents: the input and output currents. Here, we relax this standard assumption and consider the question of the efficiency fluctuations for a machine involving three different currents, first in full generality and second for two different examples. Since the third current may not be measurable and/or may decrease the machine efficiency, our motivation is to study the effect of unknown losses in small machines.
\end{abstract}

\maketitle

\newpage

\section{Introduction}

Machines use a spontaneous current to generate another one flowing against a conjugate thermodynamic force. Most machines operate on the macroscopic scale, i.e., in the thermodynamic limit, and for this reason are modeled by deterministic equations. However it is possible nowadays to build microscale machines that are strongly influenced by thermal fluctuations. Hence these microscale machines are modeled by probabilistic equations, e.g., master equation or Fokker-Planck equation~\cite{VanKampen2007_vol, Book_Risken1989, Gardiner2004_vol}. In this context, small machines are ruled by the laws of stochastic thermodynamics~\cite{Broeck2013_vol, Mou1986_vol84, VandenBroeck2014_vol418, Seifert2012_vol75}: All the thermodynamic quantities, such as heat, work, or entropy production become random variables~\cite{Jarzynski1997_vol78, Sekimoto1998_vol130, Book_Sekimoto2010, Seifert2005_vol95}. More precisely, they become functionals of the random trajectory of states visited by the machine \cite{VandenBroeck2014_vol418}.

Stochastic thermodynamics improves the weakly irreversible thermodynamics in two ways~\cite{Book_Prigogine1955, Callen1985_vol, Nicolis1977_vol}: It describes systems that are both arbitrarily irreversible and stochastic. In the last two decades, this theory has permitted revisiting the first and second laws of thermodynamics. 
Whereas the first law essentially states the conservation of energy along a unique trajectory followed by the system, the second law arises from the fluctuation theorem, i.e., a symmetry of the probability of the entropy production~\cite{Bochkov1981_vol106a, Evans1994_vol50, Gallavotti1995_vol74, Kurchan1998_vol31, Crooks2000_vol61}. This theorem ensures that the mean entropy production is positive. Since entropy production is a linear combination of the physical currents, the fluctuation theorem also affects the probability of currents~\cite{BulnesCuetara2014_vol} and, in this way, constrains the properties of small thermodynamic machines~\cite{Sinitsyn2011_vol44, Campisi2014_vol47, Campisi2015_vol17}.

Using the fluctuation theorem, the shape of the probability of the stochastic efficiency of a small machine, defined as the ratio of the stochastic output and input currents, has been predicted recently~\cite{Verley2014_vol5, Verley2014_vol90, Polettini2015_vol114, Gingrich2014_vol16,Proesmans2015_vol17,Proesmans2015_vol92}. For instance, it was shown that the probability of efficiency displays fat tails leading to a diverging first moment of the efficiency \cite{Gingrich2014_vol16,Proesmans2015_vol109}. Furthermore, the macroscopic efficiency given by the ratio of the mean currents is the most probable efficiency. On the opposite side, for stationary machines or machines operating under time-symmetric driving, the reversible efficiency is the least likely, i.e. it corresponds to a local minimum of the efficiency probability on a range that increases with the observation time of the input and output currents. 

Interestingly, some of these features have been observed experimentally by Martinez \emph{et al} for a Carnot engine on the colloidal scale~\cite{Martinez2015_vol}. In that work, the measurement of all the contributions to the entropy production turns out to be relatively complex at the fluctuating level. In general for real thermodynamic nanodevices, the difficulty is twofold: First the machine may have multiple input or output currents, and second some of these currents may not be measurable. Hence, our aim in this paper is to study the efficiency statistics when more than two currents are present, say three for simplicity. This allows in a second step to investigate efficiency fluctuations when additional currents are present but ignored.

In this context, we have in mind that the additional current models extra fueling or unknown losses in the form of work, heat flow, matter current, etc. Examples of such multiterminal machines can be found in Refs.~\cite{Mazza2014_vol16,Brandner2013_vol15,Jiang2012_vol85,Jacquet2009_vol134,Entin-Wohlman2010_vol82}. For the sake of generality, we name ``process'' any time dependent random variable yielding to a time extensive contribution in the total entropy production. A process associated with a positive (respectively, negative) entropy production on average corresponds to an input (respectively, output) process. With three processes, two types of machines exist: machines with losses (one input process and two output processes) and machines with extra fueling (two input processes and one output process). From the three processes, we can introduce two entropy production ratios. One of these ratios is interpreted as an efficiency (first output over first input), whereas the second ratio depends on the type of machine. It is either a loss factor (second output over first input) or a fueling factor (second input over first input). We notice that a loss factor could also be named an efficiency since entropy is consumed by the second output process to achieve a different task with respect to the first output process. The choice between these two names depends on the degree of usefulness of the second output process \footnote{For instance, a refrigerator can use work to cool two cold thermostat by transferring heat to a hot thermostat. In this case, it is possible to define two efficiencies, one for each cold thermostat being cooled. If our objective is to cool only one of them, then using work to cool the second one is useless and may be interpreted as a leak of cooling power.}. In this paper, we focus on studying efficiency fluctuations for a machine with losses and call generically a ratio of two entropy productions an efficiency.

On this basis, after a short thermodynamic description of a machine involving three processes in Sec.~\ref{Thermo3Processes}, we use the large deviation theory to characterize the long time statistics of the pair of stochastic efficiencies of the machine in Sec.~\ref{EffFluctFull}. The aforementioned results for machines with only two processes are recovered and extended in this section. Our main result is that the least likely efficiencies are linearly constrained one to another. For machines with time reversal symmetry, this constraint states that the least likely efficiencies sum to one. In Sec.~\ref{EffFluctCG}, we consider the case where a third process exists but is ignored in the theoretical description of the machine. We show that the statistics of the remaining efficiency has the same structure as the one predicted for a machine with only two processes. For stationary machines or machines operating under time-symmetric driving, a central difference is that the least likely efficiency is translated with respect to the reversible efficiency. It corresponds to the most reversible efficiency that is achievable considering that the third process evolves typically. We illustrate our results on two solvable models in Sec.~\ref{application}, first on a machine with a Gaussian statistics for the entropy productions [an assumption generically satisfied in the close-to-equilibrium limit], and second on a photoelectric device made of two single level quantum dots \cite{Cleuren2012_vol108,Rutten2009_vol80}.

\section{Thermodynamics of an engine with three processes \label{Thermo3Processes}}

We consider the generic case of a machine described by three thermodynamic forces  $A_1$, $A_2$, and $A_3$, and three time-integrated currents $\mathcal{J}_1$, $\mathcal{J}_2$, and $\mathcal{J}_3$. We define the currents as positive when flowing toward the machine.
The stochastic entropy production along a trajectory of duration $t$ is $\Sigma = \Sigma_1 +\Sigma_2+\Sigma_3+\Delta S$ with $\Sigma_i= \mathcal{J}_i A_i$ as the stochastic entropy production of process $i = 1,2,3$. The stochastic entropy change of the machine itself is $\Delta S$. We consider only small machines with finite state space for which the entropy change is negligible with respect to the entropy productions over a long time $t$. In this case, the total entropy production rate is given by
\begin{equation}
\sigma = \f{\Sigma}{t} = \sigma_1+\sigma_2+\sigma_3,
\end{equation}
with $\sigma_i = \Sigma_i/t$ as the entropy production rate associated with process $i = 1,2,3$. The mean value of a stochastic variable is denoted by brackets $\langle \dots \rangle$ and corresponds to averaging over all the trajectories.

In general, a device operating as a machine (on average) uses a
fueling process (the input) flowing in the direction
of its corresponding forces and therefore $\moyenne{\sigma_1} > 0 $ (e.g., heat flowing
down a temperature gradient)
in order to power a second process (the output) flowing against the direction of its
corresponding forces $\moyenne{\sigma_2} <0 $ (e.g., a particle flowing up a
chemical potential gradient). Our third process will either flow spontaneously $\moyenne{\sigma_3} >0 $,  and the machine will have two input processes, or in the opposite direction $\moyenne{\sigma_3} <0 $, and the machine will have two output processes. We define the stochastic efficiencies $\eta_1$, $\eta_2$,  and $\eta_3$ by
\begin{equation}\label{DefEta}
\eta_1= -\f{\sigma_1}{\sigma_1}=-1, \quad \eta_2 = -\f{\sigma_2}{\sigma_1} \quad \text{and} \quad \eta_3 = -\f{\sigma_3}{\sigma_1},
\end{equation}
where $\eta_1$ has been introduced by convention. The dimensionless $\eta_i$'s are ``type II" efficiencies. The ``type I" efficiencies involve the ratio of the currents and are easily recovered from the ``type II" efficiencies using the thermodynamic forces \cite{Book_Bejan2006}. The most probable values of $\eta_2$ and $\eta_3$ converge in the long time limit to the macroscopic efficiencies $\bar{\eta}_2$ and $\bar{\eta}_3$ defined by
\begin{equation}\label{EqThermoEff}
\bar{\eta}_2 = -\f{\moyenne{\sigma_2}}{\moyenne{\sigma_1}} \qquad \text{and} \qquad \bar{\eta}_3 = -\f{\moyenne{\sigma_3}}{\moyenne{\sigma_1}},
\end{equation}
which are the conventional thermodynamic efficiencies.
Since the second law imposes $\moyenne{\sigma} \geqslant 0$, we have the following constraint on the macroscopic efficiencies:
\begin{equation}
\bar{\eta}_2+\bar{\eta}_3\leqslant 1, \label{EqSecondLaw}
\end{equation}
that is reminiscent of the Carnot bound for machines with two processes and a unique efficiency. We remark here that the third process may model losses since it decreases the upper bound of the efficiency $\bar \eta_2 \leq 1-\bar \eta_3$.

\section{Efficiency statistics of a machine with three processes: General approach \label{EffFluctFull}}

Below, we study the fluctuations of the efficiencies $(\eta_2,\eta_3)$ considering that the statistics of all the entropy productions $(\sigma_1,\sigma_2,\sigma_3)$ is accessible.

\subsection{Definition of the large deviation function of the efficiencies}

The large deviation theory provides a formal framework to describe the probability of time integrated observables in the long-time limit \cite{Touchette2009_vol478,AmirDembo1998_vol}. It allows for characterizing quantitatively the exponential convergence of a probability toward a Dirac distribution centered on the mean value of the random variable studied. This rate of convergence of the probability is called a large deviation function (LDF) or a rate function. We denote by $P_t(\sigma_1,\sigma_2,\sigma_3)$ the probability density of the entropy production rates $\sigma_1,\sigma_2,\sigma_3$ after a time $t$. Assuming that a large deviation principle holds, this probability density is asymptotically given at long times by 
\begin{equation}\label{LDFSig}
P_t(\sigma_1,\sigma_2,\sigma_3) \asymp \exp{\{-tI(\sigma_1,\sigma_2,\sigma_3)\}}.
\end{equation}
The sign $\asymp$ indicates that the $o(t)$ terms in the exponent are ignored as is usual in large deviation theory \cite{Touchette2009_vol478}. By construction, the LDF $I(\sigma_1,\sigma_2,\sigma_3)$ is non-negative and assumed to be convex. Its minimum value zero is reached at the point $\left(\moyenne{\sigma_1},\moyenne{\sigma_2},\moyenne{\sigma_3}\right)$. Following Ref.~\cite{Verley2014_vol90}, we obtain the LDF of the efficiencies from the LDF of the entropy productions. The joint probability density at time $t$ to observe efficiencies $\eta_2$ and $\eta_3$ is given by
\begin{multline} \label{asympProba}
P_t(\eta_2,\eta_3) = \int{} \dd \sigma_1 \dd \sigma_2 \dd \sigma_3 P_t(\sigma_1,\sigma_2,\sigma_3) \\
\times \delta\left(\eta_2 + \f{\sigma_2}{\sigma_1}\right)\delta\left(\eta_3 + \f{\sigma_3}{\sigma_1}\right).
\end{multline}
Using Eq. (\ref{LDFSig}) in Eq.~(\ref{asympProba}) and the saddle point method to compute the integral, we find for long times,
\begin{equation}
P_t(\eta_2,\eta_3) \asymp \exp{\{-tJ(\eta_2,\eta_3)\}},
\end{equation}
where
\begin{equation}\label{MinJ}
J(\eta_2,\eta_3) = \min_{\sigma_{1}} I(\sigma_1,-\eta_2\sigma_1,-\eta_3\sigma_1)\}.
\end{equation}
From this, we deduce that $J$ is a non-negative and bounded function for all $\eta_2,\eta_3$
\begin{equation}\label{MaxJ}
 \quad 0 \leqslant J(\eta_2,\eta_3) \leqslant I(0,0,0).
\end{equation}
The efficiency LDF also follows from the cumulant generating function (CGF), 
\begin{equation}
\phi(\gamma_1,\gamma_2,\gamma_3) = \lim_{t \rightarrow \infty} \frac{1}{t} \ln \left \langle e^{ t(  \gamma_1 \sigma_1+\gamma_2 \sigma_2+\gamma_3 \sigma_3 ) } \right \rangle,
\end{equation}
of the entropy productions \cite{Verley2014_vol90}. Indeed, when $I$ is convex, $\phi$ and $I$ are conjugated by the Legendre transform,
\begin{equation}
I(\sigma_{1},\sigma_{2},\sigma_{3})=\max_{\gamma_{1},\gamma_{2},\gamma_{3}}\Big\{\sum_{i=1}^{3}\gamma_{i}\sigma_{i}-\phi(\gamma_{1},\gamma_{2},\gamma_{3})\Big\}.
\end{equation}
From this duality, we prove in Appendix~\ref{AppCGF} that
\begin{equation}\label{MinCGF}
J(\eta_2,\eta_3)=-\min_{\gamma_{2},\gamma_{3}} \phi(\gamma_{2}\eta_2+\gamma_{3}\eta_3,\gamma_{2},\gamma_{3}).
\end{equation}
This formula is of particular interest since CGFs are more convenient to compute in practice.

\subsection{Extrema of the LDF}\label{LLE}

In this section we look for the specific features of the various extrema of the efficiency LDF $J$. We first show that the location of the maxima follows from a linear constraint on the efficiencies, second that $J$ has a unique global minimum, and third that no other extremum exists at finite values of the efficiency. All these features are illustrated in Sec.~\ref{numeric} on two specific models.

\subsubsection{Maximum of the efficiency LDF}
\label{Maximum}
We look for the location of the maxima of $J$. Since we have $J(\eta_2,\eta_3) \leqslant I(0,0,0) $, if there exists at least one couple $(\eta_2^*,\eta^*_3)$ satisfying
\begin{equation}
J(\eta_2^*,\eta_3^*)=I(0,0,0), \label{maxConstraint}
\end{equation}
then $(\eta_2^*,\eta^*_3)$ is the position of a maximum. We show in Appendix~\ref{AppLLE} that, in fact, an ensemble of efficiencies verifies Eq.~(\ref{maxConstraint}). This ensemble is a straight line on the plane $(\eta_2,\eta_3)$ and is given by 
\begin{equation}
\eta_2^*\Dp{I}{\sigma_2}\bigg|_0\left(\Dp{I}{\sigma_1}\bigg|_0\right)^{-1}+\eta_3^*\Dp{I}{\sigma_3}\bigg|_0\left(\Dp{I}{\sigma_1}\bigg|_0\right)^{-1}=1, \label{EqMaxLocation}
\end{equation}
where the subscript $0$ indicates evaluation in the origin. More specifically, in the case of a machine operating at steady state or subject to time-symmetric driving cycles, we
retrieve thanks to the fluctuation theorem that $\left. \partial I/ \partial \sigma_i \right|_0 = -1/2 $,
yielding 
\begin{equation}\label{EqRevEff}
\eta_2^*+\eta_3^*=1. 
\end{equation}
From Eq.~(\ref{EqSecondLaw}) we see that the efficiencies satisfying Eq.~(\ref{EqRevEff}) correspond to efficiencies obtained along the reversible trajectories (even though the system is out of equilibrium). The unique, reversible, and least likely efficiency of an engine with two processes is replaced, for an engine with three processes, by a couple of reversible efficiencies, one of arbitrary value and the other one following from Eq.~(\ref{EqMaxLocation}).

\subsubsection{Global minimum of the efficiency LDF}

\label{Minimum}

Assuming the convexity and no constant region, $I$ has a unique minimum at $\left(\moyenne{\sigma_1},\moyenne{\sigma_2},\moyenne{\sigma_3}\right)$. The efficiency LDF $J$ vanishes at the macroscopic efficiencies $(\bar{\eta}_2,\bar{\eta}_3)$ given by Eq.~(\ref{EqThermoEff}), 
\begin{equation}\label{EqLocationMinimun}
J(\bar{\eta}_2,\bar{\eta}_3) = \min_{\sigma_{1}} I\left(\sigma_1,\sigma_1 \f{\moyenne{\sigma_2}}{\moyenne{\sigma_1}},\sigma_1 \f{\moyenne{\sigma_3}}{\moyenne{\sigma_1}} \right) = 0,
\end{equation}
where the minimum is reached for $\sigma_1 = \moyenne{\sigma_1}$. Since $J$ is a non-negative function, $(\bar{\eta}_2,\bar{\eta}_3)$ is
a global minimum.

If $I$ has a constant region,from the convexity of $I$, it is necessarily a region around
$\left(\moyenne{\sigma_1},\moyenne{\sigma_2},\moyenne{\sigma_3}\right)$ where the LDF of entropy production vanishes. In this
case, the minimum of $J$ is not unique, but is a domain including $(\bar \eta_2,\bar \eta_3)$.

\subsubsection{Asymptotic behavior of the efficiency LDF}
\label{Asymptotic}
Let us now verify that $J$ has no other extremum than $(\bar{\eta}_2,\bar{\eta}_3)$ and $(\eta_2^*,\eta^*_3)$. To do so, we look for the zeros of the partial derivatives of $J$ with respect to $\eta_2$ and $\eta_3$, 
\begin{equation}\label{PartialDiff}
\Dp{J}{\eta_2}(\eta_2,\eta_3) = 0 \quad \mbox{ and } \quad \Dp{J}{\eta_3}(\eta_2,\eta_3) = 0.
\end{equation}
Since $J$ follows from a minimization on $\sigma_1$, see Eq.~(\ref{MinJ}), we introduce the function $\tilde{\sigma_1}(\eta_2,\eta_3)$ as the solution of 
\begin{equation}\label{S0}
0 = \f{\dd}{\dd \tilde{\sigma_1}}\left[I(\tilde{\sigma_1},-\eta_2 \tilde{\sigma_1}, - \eta_3 \tilde{\sigma_1} )\right] = \Dp{I}{\sigma_1}  - \eta_2\Dp{I}{\sigma_2}
-\eta_3 \Dp{I}{\sigma_3},
\end{equation}
with all partial derivatives evaluated in $(\tilde{\sigma_1}, -\eta_2 \tilde{\sigma_1}, - \eta_3 \tilde{\sigma_1} )$. This allows to write the efficiency LDF as
\begin{equation}\label{Jsigtilde}
J(\eta_2,\eta_3) = I(\tilde{\sigma_1}(\eta_2,\eta_3), -\eta_2 \tilde{\sigma_1}(\eta_2,\eta_3), - \eta_3 \tilde{\sigma_1}(\eta_2,\eta_3)) .
\end{equation}
From this equation, the partial derivative of $J$ may be written as
\begin{equation}\label{DiffEnd}
\Dp{J}{\eta_2}(\eta_2,\eta_3) = \Dp{\tilde{\sigma_1}}{\eta_2} \Dp{I}{\sigma_1} - \left(\eta_2 \Dp{\tilde{\sigma_1}}{\eta_2} + \tilde{\sigma_1}\right)\Dp{I}{\sigma_2} 
 - \eta_3 \Dp{\tilde{\sigma_1}}{\eta_2} \Dp{I}{\sigma_3},
\end{equation}
where partial derivatives are still taken at $(\tilde{\sigma_1},-\eta_2 \tilde{\sigma_1}, - \eta_3 \tilde{\sigma_1} )$ with $\tilde \sigma_1 = \tilde \sigma_1 (\eta_2,\eta_3)$.
From Eqs.~(\ref{S0}) and (\ref{DiffEnd}), it is possible to rewrite Eq.~(\ref{PartialDiff}) as
\begin{eqnarray}
\tilde{\sigma_1} \Dp{I}{\sigma_2} (\tilde{\sigma_1},-\eta_2 \tilde{\sigma_1}, - \eta_3 \tilde{\sigma_1} )  &=& 0, \\
 \tilde{\sigma_1} \Dp{I}{\sigma_3} (\tilde{\sigma_1},-\eta_2 \tilde{\sigma_1}, - \eta_3 \tilde{\sigma_1} ) &=& 0.
\end{eqnarray}
We distinguish now two different cases: First, the partial derivatives of $I$ may vanish, and we recover the minimum of $J$ studied in Sec.~\ref{Minimum}; second, the function $\tilde \sigma_1(\eta_2,\eta_3)$ vanishes. In the latter case, we look for $(\tilde \eta_2, \tilde \eta_3)$ such that $\tilde \sigma_1(\tilde \eta_2, \tilde \eta_3) = 0$. In this view, we evaluate Eq.(\ref{Jsigtilde}) at $(\tilde \eta_2, \tilde \eta_3)$ yielding, if $\tilde \eta_2$ and $\tilde \eta_3$ are finite,
\begin{equation}
J(\tilde \eta_2, \tilde \eta_3) = I(0,0,0),
\end{equation}
such that we retrieve the extrema $(\tilde \eta_2, \tilde \eta_3) \in (\eta^*_2,\eta^*_3)$ of Sec.~\ref{Maximum}. Alternatively, if one of the efficiencies, say $\eta_2$ for instance, is infinite, Eq.~(\ref{Jsigtilde}) becomes
\begin{multline}
\lim\limits_{\eta_2 \to \pm \infty} J(\eta_2,\eta_3) \\ = \lim\limits_{\eta_2 \to \pm \infty} I\left(\tilde \sigma_1(\eta_2,\eta_3),-\eta_2 \tilde \sigma_1(\eta_2,\eta_3), -\eta_3 \tilde \sigma_1(\eta_2,\eta_3)\right) \\ 
\leqslant I(0,0,0).
\end{multline}
From the last inequality and the convexity of $I$ we conclude that $ \eta_2 \tilde \sigma_1(\eta_2,\eta_3) $ stays finite when $\eta_2 \to \pm \infty$, and necessarily,
\begin{equation}
\lim\limits_{\eta_2 \to \pm \infty} \tilde \sigma_1(\eta_2,\eta_3 ) = 0.
\end{equation}
The derivative of $J$ vanishes at infinite efficiencies, and the efficiency LDF converges to a finite value at large efficiencies since $J$ is bounded. Moreover because the limit $\lim\limits_{\eta_2 \to \pm \infty} \eta_2 \tilde \sigma_1(\eta_2,\eta_3)$ is a constant independent of $\eta_3$ it follows that the limit $\lim\limits_{\eta_2 \to \pm \infty} J(\eta_2,\eta_3)$ is also independent of $\eta_3$ if $\eta_3$ remains finite. The same arguments hold when taking the limit $\eta_3 \to \pm \infty$ keeping $\eta_2$ finite. In the end, we have recovered all the extrema at finite values of the efficiencies and shown that the two partial derivatives of $J$ vanish at large efficiencies.

\section{Efficiency statistics of a machine with three processes: Forgetting the third process\label{EffFluctCG}}


We now study the fluctuations of the efficiency $\eta_2$ without taking into account the statistics on the third process. This may correspond to an experimental setup for which the third current exists but cannot be measured. In this case, we consider that $\eta_3$ (or equivalently $\sigma_3$) always takes the typical value associated with some given efficiency $\eta_2$: This leads to contracting the LDF $J(\eta_2,\eta_3)$ on $\eta_3$. We analyze in this section the general shape of the contracted LDF and study its extrema. 

The contracted LDF is by definition
\begin{equation}\label{minJc}
J_{\ct}(\eta_2) = \min_{\eta_3} J(\eta_2,\eta_3) =  \min_{\sigma_1} I_\ct(\sigma_1,-\eta_2\sigma_1),
\end{equation}
with 
\begin{equation}\label{MinimizationI'}
I_\ct(\sigma_1,\sigma_2) =  \min_{\eta_3}I(\sigma_1,\sigma_2,-\eta_3\sigma_1) =  \min_{\sigma_3}I(\sigma_1,\sigma_2,\sigma_3) .
\end{equation}
As in the previous case, cf. Appendix~\ref{AppCGF}, we can express the contracted efficiency LDF in terms of the CGF,
\begin{equation}
J_{\ct}(\eta_2) = -\min_{\gamma_2} \phi(\gamma_2 \eta_2, \gamma_2,0).
\end{equation}
We now determine some properties of this contracted LDF. From (\ref{minJc}), we have for all $\eta_2$
\begin{equation}
0 \leqslant J_{\ct}(\eta_2) \leqslant I_\ct(0,0),
\end{equation}
so $J_{\ct}$ is a non-negative bounded function. In particular, we are interested in the extrema of $J_{\ct}$. 

First, looking for the minimum, we have
\begin{equation}
J_{\ct}(\bar{\eta}_2) \leqslant J(\bar{\eta}_2,\bar{\eta}_3) = 0,
\end{equation}
so, due to the positivity of $J_{\ct}$, the efficiency $\bar{\eta}_2$ is a global minimum of $J_{\ct}$ and corresponds to the macroscopic efficiency.

Second, we look for the maximum of $J_{\ct}(\eta_2)$. We call $\eta^*_{2,\ct}$ the efficiency such that $J_{\ct}(\eta^*_{2,\ct})=I_\ct(0,0)$, and, reasoning as in Appendix \ref{AppLLE}, we have 
\begin{equation}\label{Etaetoile}
\eta^*_{2,\ct}=\left(\Dp{I_\ct}{\sigma_1}\bigg|_0\right)\left(\Dp{I_\ct}{\sigma_2}\bigg|_0\right)^{-1},
\end{equation}
Since $I_\ct$ follows from the minimization of Eq.~(\ref{MinimizationI'}) over $\sigma_3$, we introduce $\tilde{\sigma_3}(\sigma_1,\sigma_2)$ as the solution of this minimization, yielding,
\begin{equation}
I_\ct(\sigma_1,\sigma_2)= I(\sigma_1,\sigma_2,\tilde\sigma_3(\sigma_1,\sigma_2)).
\end{equation}
And next, we find
\begin{eqnarray}
\Dp{I_\ct}{\sigma_2}\bigg|_0 = \Dp{I}{\sigma_2}(0,0,\tilde{\sigma_3}(0,0)), \\
\Dp{I_\ct}{\sigma_1}\bigg|_0= \Dp{I}{\sigma_1}(0,0,\tilde{\sigma_3}(0,0)).
\end{eqnarray}
After contraction on $\sigma_3$, Eq.~(\ref{Etaetoile}) yields the least likely efficiency,
\begin{multline}
\eta^*_{2,\ct}=\left(\Dp{I}{\sigma_1}(0,0,\tilde{\sigma_3}(0,0))\right)\left(\Dp{I}{\sigma_2}(0,0,\tilde{\sigma_3}(0,0))\right)^{-1}.
\label{LLeffCT}
\end{multline}
In this equation we see that the least likely efficiency is achieved when processes $1$ and $2$ evolve reversibly whereas the third process evolves typically (with the condition that the first two processes are reversible). In other words, at the least likely efficiency, the system chooses the most probable trajectories compatible with the reversibility of the first two processes. Since in the general case $I_\ct$ will not satisfy a fluctuation theorem, we have no constraint on the location of the maximum of $J_\ct(\eta_2)$. If $\tilde{\sigma_3}(0,0)$ is small, a Taylor expansion of Eq.~(\ref{LLeffCT}) around $(0,0,0)$ shows that the maximum is slightly moved away from $\eta_2^*$ given by Eq.~(\ref{EqMaxLocation}) taken at $\eta_3^*=0$. But for an arbitrary value of $\tilde{\sigma_3}(0,0)$, the maximum of $J_\ct$ can be anywhere, even below $\bar{\eta}_2$. This does not contradict the second law of thermodynamics since the third process (that is ignored here) may fuel the machine as much as waste its power.

Finally, we verify the absence of another extremum of $J_\ct$ at finite efficiency. To do so, we seek as earlier the zeros of the derivative of $J_{\ct}$,
\begin{equation}
\ddf{J_{\ct}}{\eta_2} = 0.
\end{equation}
To find an expression for this derivative, we introduce the function $\tilde{\sigma}'_1(\eta_2)$ realizing the minimum in Eq.~(\ref{minJc}) such that
\begin{eqnarray}
J_{\ct}(\eta_2) &=& I_\ct(\tilde{\sigma}'_1(\eta_2),-\eta_2\tilde{\sigma}'_1(\eta_2)) \nonumber \\
                      &=& I(\tilde{\sigma}'_1(\eta_2),-\eta_2 \tilde{\sigma}'_1(\eta_2),-\tilde\sigma_3(\tilde{\sigma}'_1(\eta_2),-\eta_2 \tilde{\sigma}'_1(\eta_2))). \nonumber \\
\end{eqnarray}
The total derivative of $J_{\ct}(\eta_2)$ yields
\begin{equation}
\ddf{J_{\ct}}{\eta_2}(\eta_2) = -\tilde{\sigma}'_1(\eta_2)\Dp{I}{\sigma_2}.
\end{equation}
With arguments similar to those of Sec.~\ref{Asymptotic}, the above derivative vanishes only at the previously obtained extrema and for infinite values of efficiency. Since $J_{\ct}$ is bounded, it converges to finite values when $\eta_2 \to \pm \infty$. 

Therefore, $J_{\ct}$ has the typical shape of the efficiency LDF for two external processes \cite{Verley2014_vol90} but with a displaced maximum. An example is provided in Fig.~\ref{FigContractionFE}.

\section{Applications \label{application} }

\subsection{Close-to-equilibrium machine \label{secCE}}

Close to equilibrium, the cumulant generating function of entropy productions is generically a quadratic function,
\begin{equation}\label{GaussCGF}
\phi(\gamma_{1},\gamma_{2},\gamma_{3}) = \sum_{i,j=1}^{3} C_{i,j} \gamma_i\gamma_j + \sum_{i=1}^{3} \gamma_i \moyenne{\sigma_i},
\end{equation}
with $C_{i,j}$ as the asymptotic covariances of the entropy productions defined by
\begin{equation}
C_{i,j} = \lim_{t\rightarrow \infty}\frac{\moyenne{\Sigma_i(t)\Sigma_j(t)}-\moyenne{\Sigma_i(t)}\moyenne{\Sigma_j(t)}}{t}.
\end{equation}
From Eqs. (\ref{MinCGF}) and (\ref{GaussCGF}) we calculate the efficiency LDF,
\begin{multline}\label{Jgauss}
J(\eta_2,\eta_3) \\ = \f{ \Sum{i,j=2}{3} \left( \moyenne{\sigma_{i}}+\eta_{i} \moyenne{\sigma_1} \right) M_{5-i,5-j} \left( \moyenne{\sigma_{j}} +\eta_{j} \moyenne{\sigma_1} \right)}
{\Sum{s,s'}{} \epsilon(s)\epsilon(s') \eta_{s(1)}\eta_{s'(1)}C_{s(2),s'(2)}C_{s(3),s'(3)}}
\end{multline}
where $s$ denotes a permutation of three elements and $\epsilon(s)$ denotes its parity and
\begin{equation}
M_{i,j}= (-1)^{i+j}\left(C_{i,j}+ C_{1,i}\bar\eta_j + C_{1,j}\bar\eta_i +C_{1,1}\bar\eta_i\bar\eta_j  \right)
\end{equation}
for $i,j=2,3$.  
We can also rewrite $J(\eta_2,\eta_3)$ in a form that is convenient for generalization,
\begin{multline}\label{NEqPermut}
J(\eta_2,\eta_3) \\ = \f{\Sum{s,s'}{} \epsilon(s)\epsilon(s') \eta_{s(1)}\eta_{s'(1)}C_{s(2),s'(2)}\moyenne{\sigma_{s(3)}}\moyenne{\sigma_{s'(3)}}}{\Sum{s,s'}{} \epsilon(s)\epsilon(s') \eta_{s(1)}\eta_{s'(1)}C_{s(2),s'(2)}C_{s(3),s'(3)}}.
\end{multline}
As in Ref.~\cite{Verley2014_vol90}, the close-to-equilibrium efficiency LDF is the ratio of two quadratic forms. It vanishes as expected at the macroscopic efficiencies $(\bar{\eta}_2,\bar{\eta}_3)$. A comparison between the close-to-equilibrium case and a general calculation on efficiency LDF is provided in Sec.~\ref{numeric} for a specific model.

Furthermore, from linear response theory, the mean entropy production rates are connected to the asymptotic covariances of entropy production as follows:
\begin{equation}
\moyenne{\sigma_i} = \f{1}{2}\Sum{j=1}{3} C_{i,j}
\end{equation} 
Then, Eq. (\ref{NEqPermut}) may be rewritten using only the coefficient $C_{i,j}$,
\begin{multline}\label{NEqPermutC}
J(\eta_2,\eta_3) \\ = \f{\displaystyle \Sum{s,s'}{} \sum_{i,j=1}^3 \epsilon(s)\epsilon(s') \eta_{s(1)}\eta_{s'(1)}C_{s(2),s'(2)}C_{i,s(3)}C_{j,s'(3)}}{2\Sum{s,s'}{}\epsilon(s)\epsilon(s') \eta_{s(1)}\eta_{s'(1)}C_{s(2),s'(2)}C_{s(3),s'(3)}}.
\end{multline}
Since the asymptotic covariances are proportional to the response coefficient of the machine, the close-to-equilibrium efficiency LDF is completely known from the response property of the machine.

From this LDF for the two efficiencies we now explicitly compute $J_\ct$. After the contraction on the efficiency $\eta_3$, we retrieve the functional form of the efficiency LDF for a machine with two processes~\cite{Verley2014_vol90},
\begin{equation}\label{Jcgauss}
J_{\ct}(\eta_2)=\f{1}{2}\f{(\eta_2\moyenne{\sigma_1}+\moyenne{\sigma_2})^2}{(\eta_2)^2 C_{1,1}+2\eta_2 C_{1,2}+C_{2,2}},
\end{equation}
keeping in mind that we have now $\moyenne{\sigma_i} = \sum_{j=1}^3 C_{i,j}/2$ and not $\moyenne{\sigma_i} = \sum_{j=1}^2 C_{i,j}/2$ as in Ref.~\cite{Verley2014_vol90}. The maximum is no longer at $\eta_2=1$ but at $\eta_2=\eta^*_{2,\ct}$ with
\begin{equation}
\eta^*_{2,\ct} = \f{C_{2,2}C_{1,1}+C_{1,3}C_{2,2}-C_{1,2}^2-C_{1,2}C_{2,3}}{C_{2,2}C_{1,1}+C_{2,3}C_{1,1}-C_{1,2}^2-C_{1,2}C_{1,3}}.
\end{equation}
As expected, when $C_{1,3}$ and $C_{2,3}$ vanish, $\eta^*_{2,\ct}=1$: When the third process decouples from the others, we retrieve the least likely efficiency of a machine with only two processes.

\subsection{Photoelectric device \label{numeric}}

\begin{figure*}
\centering
\includegraphics[width=8cm,bb=0 184 497 576]{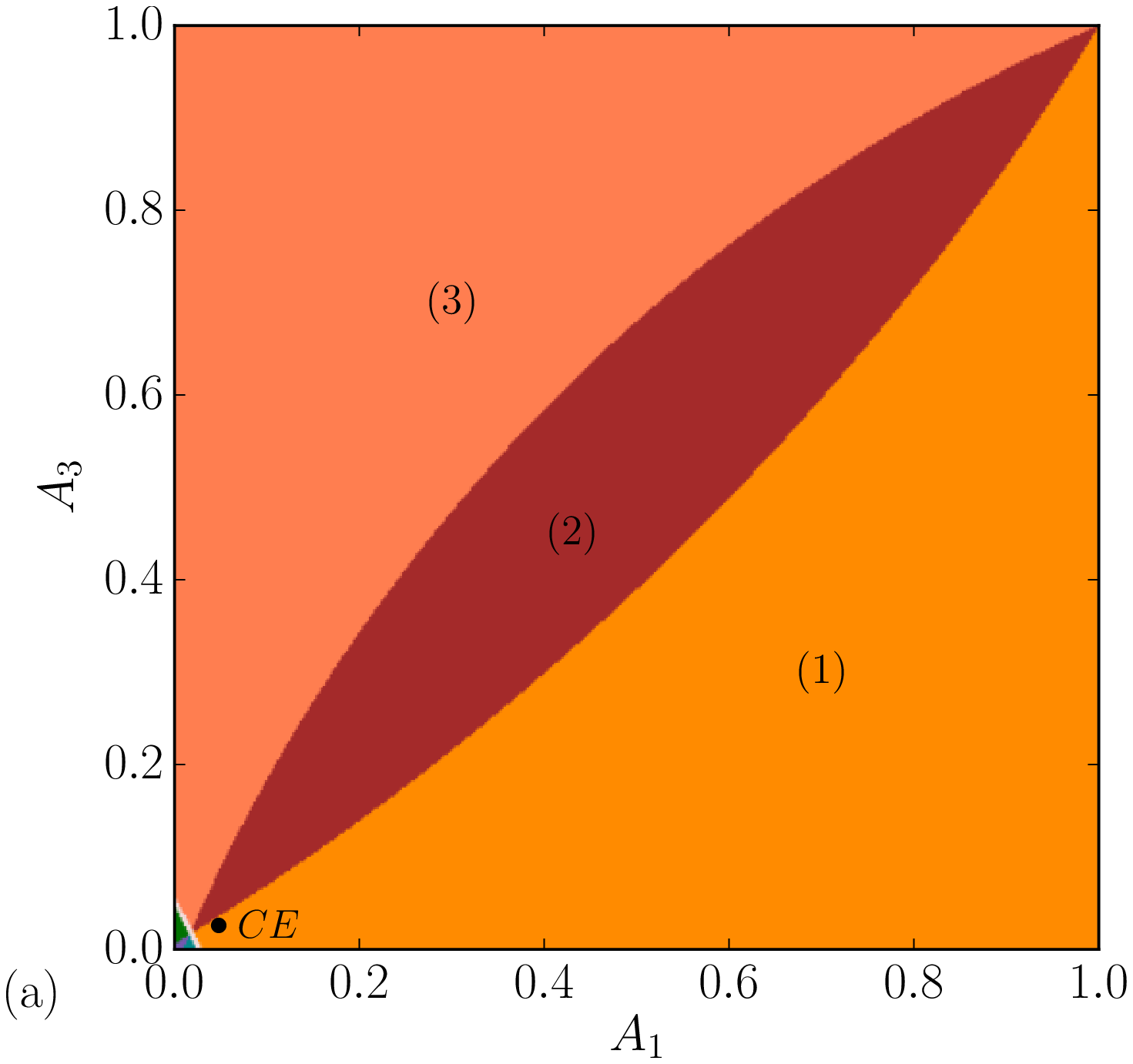}  \includegraphics[width=8cm,bb=0 184 497 576]{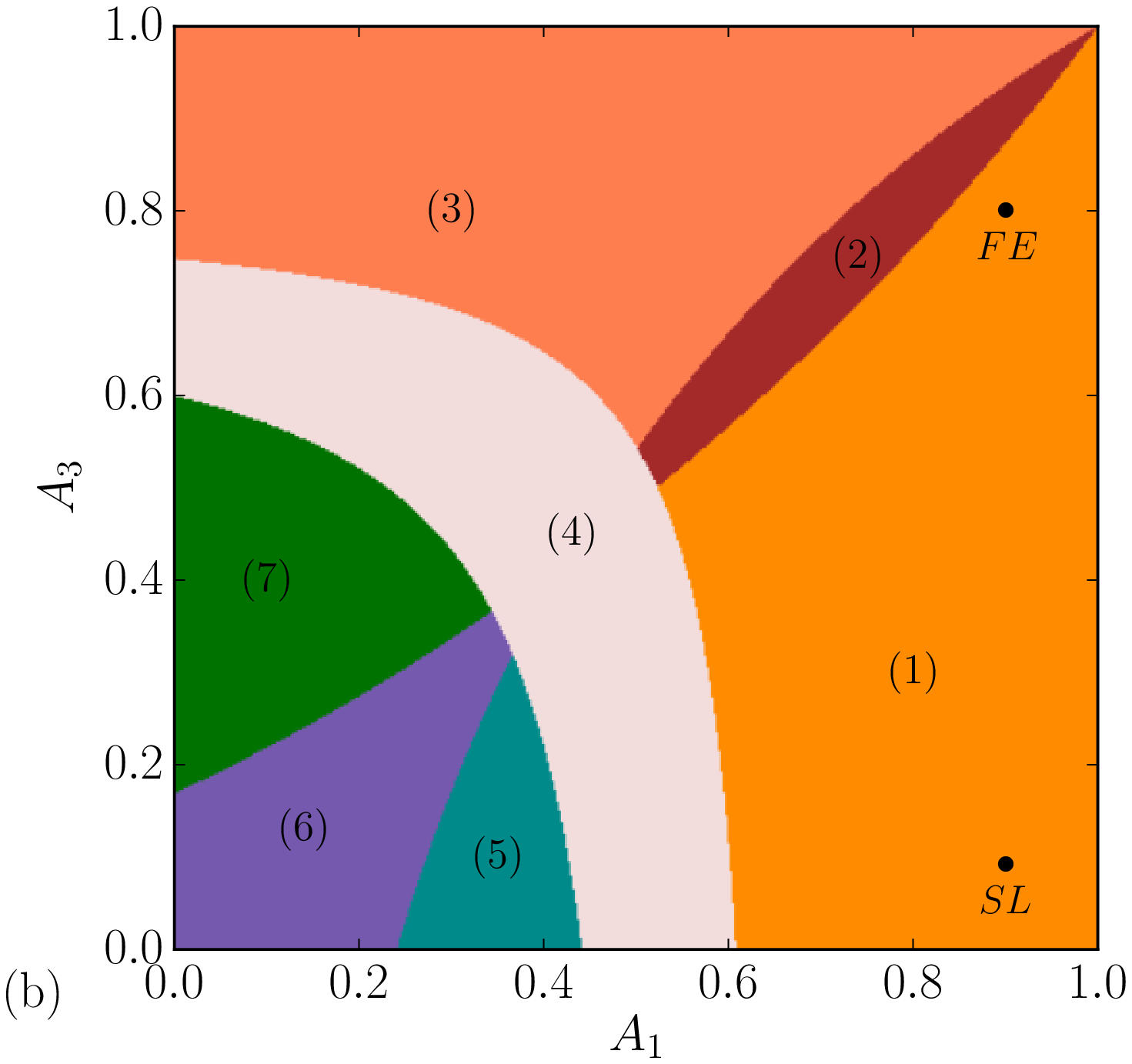}
\caption{Diagram representing the various operating modes of the photoelectric cell as a function of the affinities $A_1$ and $A_3$ for (a) a small chemical potential difference $\Delta \mu =0.01$ and (b) a large chemical potential difference $\Delta \mu = 1$. The black dots correspond to the three studied cases: the close-to-equilibrium case ``CE'', the small loss case ``SL'', and the far from equilibrium case ``FE''.
(1)-(3): heat engine, for each label $\moyenne{w} < 0 $, $\moyenne{q_c} < 0 $, and more specifically 
(1) $\moyenne{q_h} > 0 > \moyenne{q_m}$,  
(2) $\moyenne{q_h} > 0 $, $\moyenne{q_m} > 0 $, 
(3) $\moyenne{q_h} < 0 < \moyenne{q_m} $, 
(4): dud engine,  $\moyenne{w} > 0 $, $\moyenne{q_c} < 0 $.
(5)-(7): refrigerator and heat pump, for each label $\moyenne{w} > 0 $, $\moyenne{q_c} > 0 $, and more specifically 
(5) $\moyenne{q_h} > 0 > \moyenne{q_m}$, 
(6) $\moyenne{q_h} < 0 $, $\moyenne{q_m} < 0 $, and
(7) $\moyenne{q_h} < 0 < \moyenne{q_m}$. 
Parameters for the machine are $E_r=2.5$, $E_l=0.5$ (energies are in units of $k_{B}T_c$), $\Gamma_c=1$, $\Gamma_m=5$, and $\Gamma_h=\Gamma_l=\Gamma_r=10$, and more specifically in the CE case:
$T_c=1$, $\mu_l=1$, $\mu_r=1.035$, $T_m=1.025$, and $T_h=1.05$
; in the FE case:
$\mu_l=1$, $\mu_r=2$, $T_c=1$, $T_m=5$, and $T_h=10$ 
; and in the SL case:
$\mu_l=1$, $\mu_r=2$, $T_c=1$, $T_m=1.1$, and $T_h=10$. 
 \label{FigDiagFonct} }
\end{figure*}

We now illustrate the results of the previous sections on a model of a photoelectric device first studied in Refs.~\cite{Cleuren2012_vol108,Rutten2009_vol80}. 
\begin{figure}
\centering
\includegraphics[width=7cm]{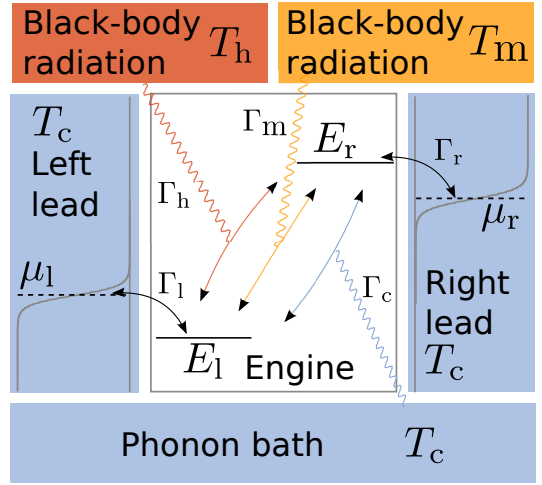}
\caption{Sketch of the photoelectric device. The device is made of two single-level quantum dots (in white)
connected to two leads (in blue) at temperature $T_c$ and at different chemical potentials $\mu_r$ and $\mu_l$. The
electron transitions between left and right quantum dots are induced either by photons from black-body radiation at
temperature $T_h$ (in red) or $T_m$ (in orange) or by phonons at temperature $T_c$ (in blue). The arrows indicate
possible electronic transitions between different energy levels, and the $\Gamma$'s represent the coupling strengths with
the reservoirs. \label{FigCell} }
\end{figure}
The device is composed of two quantum dots each with a single energy level $E_l$ and $E_r$ ($E_r > E_l$), cf. Fig.~\ref{FigCell}. It is powered by two black-body sources at temperatures $T_h$ and $T_m$ and a cold heat reservoir at temperature $T_c$. We set Boltzmann's constant to $1$ such that a temperature is homogeneous to an energy. Each quantum dot can exchange electrons with an electronic lead at temperature $T_c$, the left (right) dot being connected to the left (right) lead. Each lead is at a different voltage and is modeled by an electron reservoir at chemical potential $\mu_r > \mu_l$.  The three different states of the machine are indexed by $j=0,\,l,\,r$, corresponding to no electron in the device, one electron in the left quantum dot, and one in the right dot, respectively. The three different heat reservoirs are labeled by $\nu=c,\,m,\,h$. We introduce the rates $k_{ij}$ as the probability per unit time to jump from state $j$ to $i$. With the Fermi-Dirac distribution $f(x) = 1/(e^x+1)$ and the Bose-Einstein distribution $b(x) = 1/(e^x-1)$, these rates are written
\begin{align}
k_{0l} =& \Gamma_l\,f(\f{E_l-\mu_l}{T_c}), & k_{l0} &=\Gamma_l\,\left(1-f(\f{E_l-\mu_l}{T_c})\right), \nonumber\\
k_{0r} =& \Gamma_r\,f(\f{E_r-\mu_r}{T_c}), & k_{r0} &=\Gamma_r\,\left(1-f(\f{E_r-\mu_r}{T_c})\right),  \nonumber\\
k_{rl}^\nu =& \Gamma_\nu\,b(\f{E_r-E_l}{T_\nu}), & k_{lr}^\nu &=\Gamma_\nu\,\left(1+b(\f{E_r-E_l}{T_\nu})\right).
\end{align}
The total rate for the left to right transition is $k_{rl} = \sum_\nu k^\nu_{rl}$ and similarly for the right to left transition. The $\Gamma$'s are the different coupling strengths with the reservoirs, see Fig.~\ref{FigCell}. The machine displays various operating modes according to the parameter values as illustrated in
Fig.~\ref{FigDiagFonct}.
We consider only the heat engine case: Other operating modes follow from relabeling the various processes. Per unit time, the machine receives a heat $q_\nu = n_\nu \Delta E$ from the heat reservoir $\nu$, where $n_\nu$ is the net rate of photons (or phonons) absorbed from this reservoir and $\Delta E = E_r-E_l$. Similarly, the work delivered by the machine is $-w = n_e \Delta \mu $, where $n_e$ is the net rate of electrons transferred from the left to the right lead and $\Delta \mu = \mu_r-\mu_l$. The heat and work fluxes represent energy currents that are associated with entropy production rates $\sigma_i$ and affinities $A_i$ as follows
\begin{equation}
\sigma_1 = \left(\f{1}{T_c}-\f{1}{T_h}\right) q_h,\qquad A_1=\f{1}{T_c}-\f{1}{T_h},
\end{equation}
\begin{equation}
\sigma_2 = \f{w}{T_c} ,\qquad A_2=\f{1}{T_c},
\end{equation}
\begin{equation}
\sigma_3 = \left(\f{1}{T_c}-\f{1}{T_m}\right) q_m,\qquad A_3=\f{1}{T_c}-\f{1}{T_m}.
\end{equation}
Accordingly, for a heat engine with losses due to the third process, the two efficiencies are 
\begin{equation}
\eta_2 = \f{-\sigma_2}{\sigma_1} = \f{-w}{q_h\left(1-T_c/T_h\right)},
\end{equation}
\begin{equation}
\eta_3 = \f{-\sigma_3}{\sigma_1} = \f{-q_m\left(1-T_c/T_m\right)}{q_h\left(1-T_c/T_h\right)}.
\end{equation}

We define the generating function of the system by $g_j(t,\gamma_1,\gamma_2,\gamma_3)= \left\langle\delta_{j,j(t)}e^{\gamma_1\Sigma_1+\gamma_2\Sigma_2+\gamma_3\Sigma_3}\right\rangle$, where $\delta$ is the Kronecker symbol. This generating function evolves according to the equation \cite{Esposito2009_vol81},
\begin{widetext}
\begin{equation}
\ddf{}{t}\left( \begin{matrix}
g_0 \\ g_l \\ g_r
\end{matrix} \right) 
=  \left[ \begin{matrix}
 -k_{l0} - k_{r0}& k_{0l}& k_{0r}e^{-\gamma_2 \Delta \mu / T_c} \\[2mm]
  k_{l0} &-k_{0l}-k_{rl}^c +k_{rl}^m+k_{rl}^h &k_{lr}^c +k_{lr}^m e^{-\gamma_3\Delta E A_3}+k_{lr}^h e^{-\gamma_1\Delta E A_1} \\[2mm]
  k_{r0} e^{\gamma_2 \Delta \mu/T_c} &k_{rl}^c +k_{rl}^me^{\gamma_3\Delta E A_3}+k_{rl}^he^{\gamma_1\Delta E A_1} &-k_{0r}- k_{lr}\\
\end{matrix}\right] \left( \begin{matrix}
g_0 \\ g_l \\ g_r
\end{matrix}
\right). \label{EvoOfGenFun}
\end{equation}
\end{widetext}
For $\gamma_1=\gamma_2=\gamma_3=0$, we retrieve the master equation for the probability $g_j(t,0,0,0) = \left\langle\delta_{j,j(t)}\right\rangle$ to be in state $j$ at time $t$.

\begin{figure*}
\centering
\hspace{0.8cm}\includegraphics[width=7.2cm]{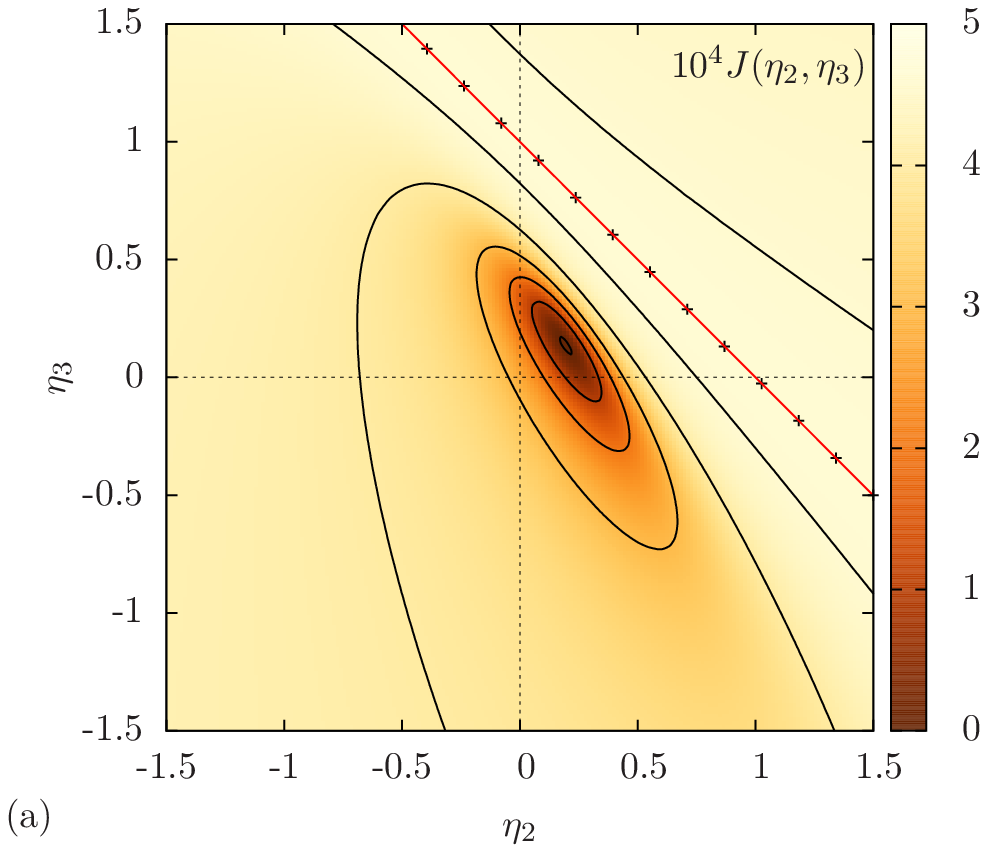}   \includegraphics[width=7.55cm]{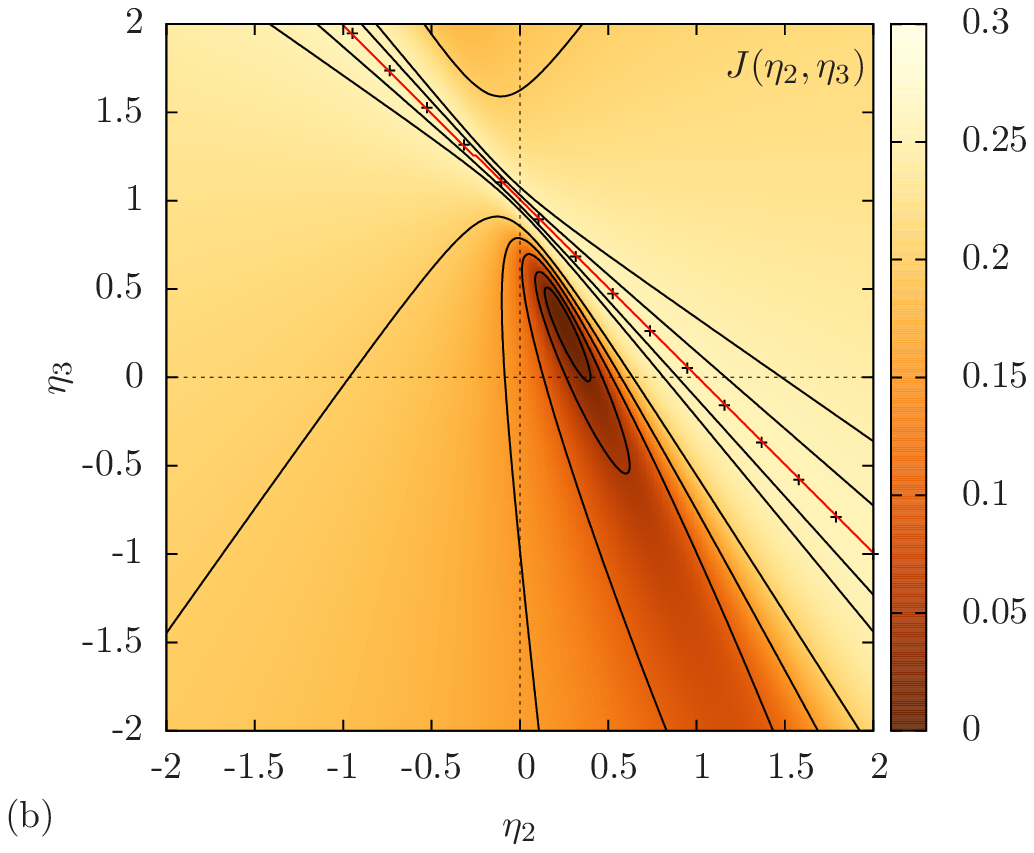}

\includegraphics[width=7.2cm]{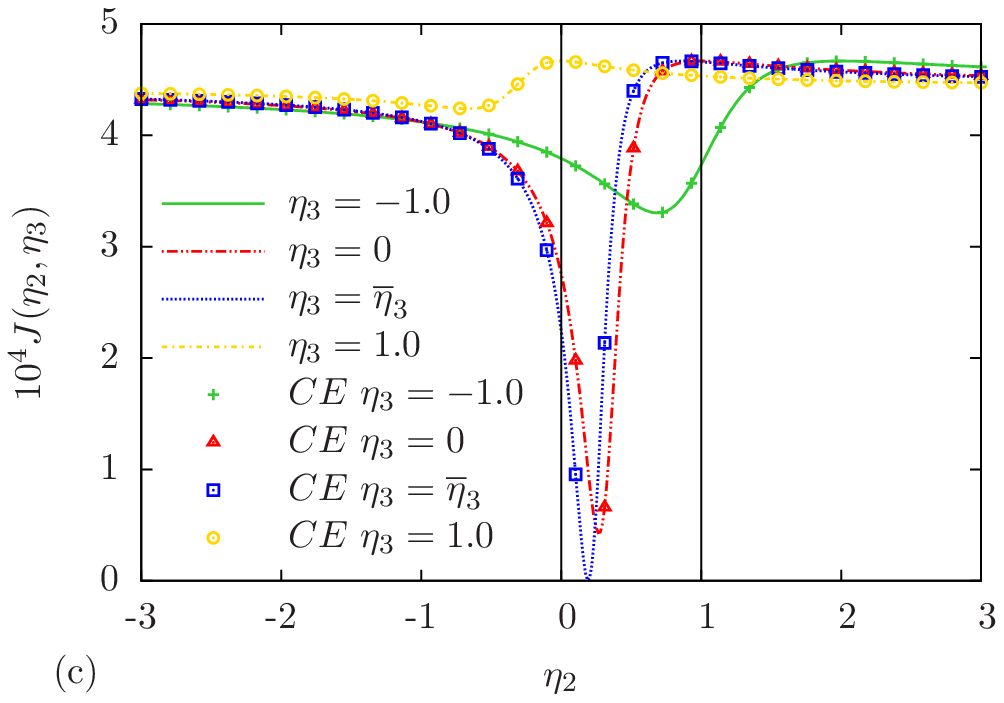}    \includegraphics[width=7.2cm]{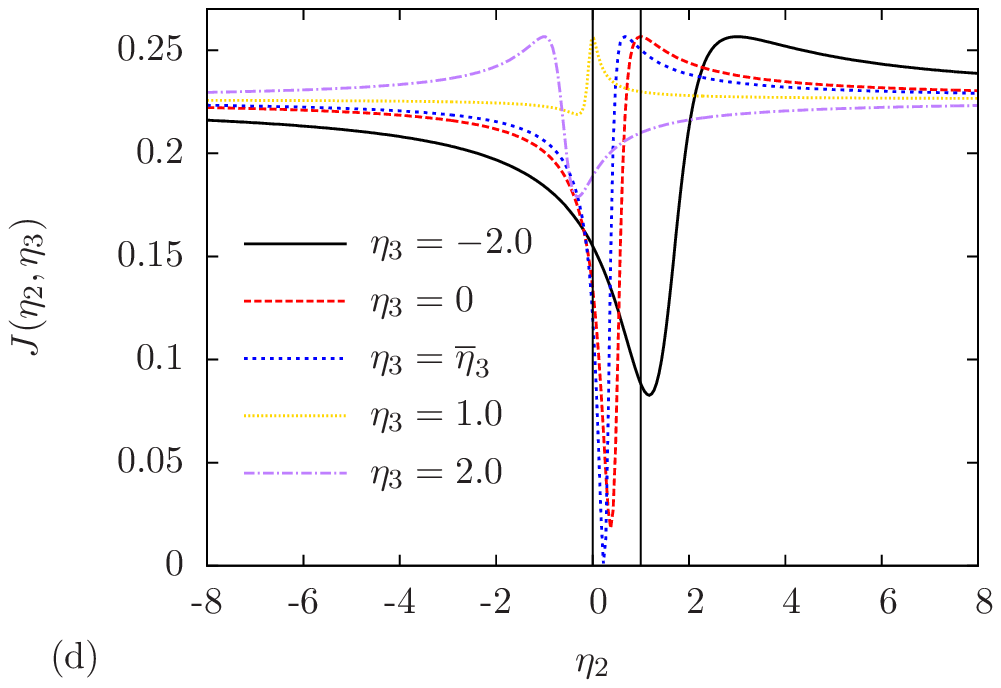}

\caption{(a) and (b) Efficiency LDF $J(\eta_2,\eta_3)$ for the photoelectric device of Fig.~\ref{FigCell} operating on average as a heat engine. The maximum of $J$ is achieved on the red contour line whereas black solid lines are contour lines for smaller $J$. The straight line of equation $\eta_2+\eta_3=1$ is shown with cross marks. (c) and (d) Crosssections of $J(\eta_2,\eta_3)$ for various $\eta_3$'s. Symbols in (c) are obtained from Eq.~(\ref{NEqPermutC}).
The figures on the left and on the right are for the CE and the FE cases, respectively, see the parameters of Fig.~\ref{FigDiagFonct}. \label{FigJ} }
\end{figure*}

Below, the fluctuations of the efficiencies $(\eta_2,\eta_3)$ are quantitatively analyzed in three different cases: a close-to-equilibrium (CE) case, a far-from-equilibrium (FE) case, and a small loss (SL) case. The parameter values in each case are summarized in the caption of Fig.~\ref{FigDiagFonct}. The efficiency statistics has been obtained first by computing numerically the highest eigenvalue of the matrix in the right hand side of Eq.~(\ref{EvoOfGenFun}) yielding the CGF $\phi$ of the various entropy production rates, and in a second step, by using Eq.~(\ref{MinCGF}) to get the efficiency LDF from $\phi$. The code is written in \textsc{Python} 3 and uses the algorithms implemented in the \textsc{Scipy} library \cite{Jones2001_vol}.

In Figs.~\ref{FigJ}(a) and~\ref{FigJ}(b) we show the efficiency LDF $J(\eta_2,\eta_3)$ in the CE and FE cases, respectively. As expected, the maximum of $J$ is located on the line $\eta_2+\eta_3=1$ corresponding to the reversible efficiencies. The minimum corresponds to the macroscopic efficiencies $(\bar \eta_2,\bar \eta_3) = (0.19,0.14)$ in the CE case and to $(\bar \eta_2,\bar \eta_3) =(0.24,0.33) $ in the FE case.

In Fig.~\ref{FigJ}(c) we verify the validity of the CE limit developed in Sec.~\ref{secCE}. The crosssections of the efficiency LDF $J$ obtained by direct numerical computation are in perfect agreement with the same crosssections but obtained from Eq.~(\ref{NEqPermutC}). In Fig.~\ref{FigJ}(d), we also show the crosssections of $J$ but in the FE case illustrating that all the fluctuations associated with a large efficiency become generically equally likely whatever the value of the other efficiency: The LDF flattens and converges to the same limit at infinity for the different crosssections. Comparing Figs.~\ref{FigJ}(c) and \ref{FigJ}(d), we remark that the time scale on which an efficiency fluctuation disappears is much longer close to equilibrium than far from equilibrium. The order of magnitude of this time scale is roughly the inverse of the maximum value of the efficiency LDF. Since this maximum is achieved for trajectories with null entropy production and from the fact that trajectories with small entropy production are more likely to appear close to equilibrium, we conclude that the efficiency fluctuations have higher probability and accordingly take more time to decay in the CE case than in the FE case.

\begin{figure*}[t]
\centering
\includegraphics[width=8cm]{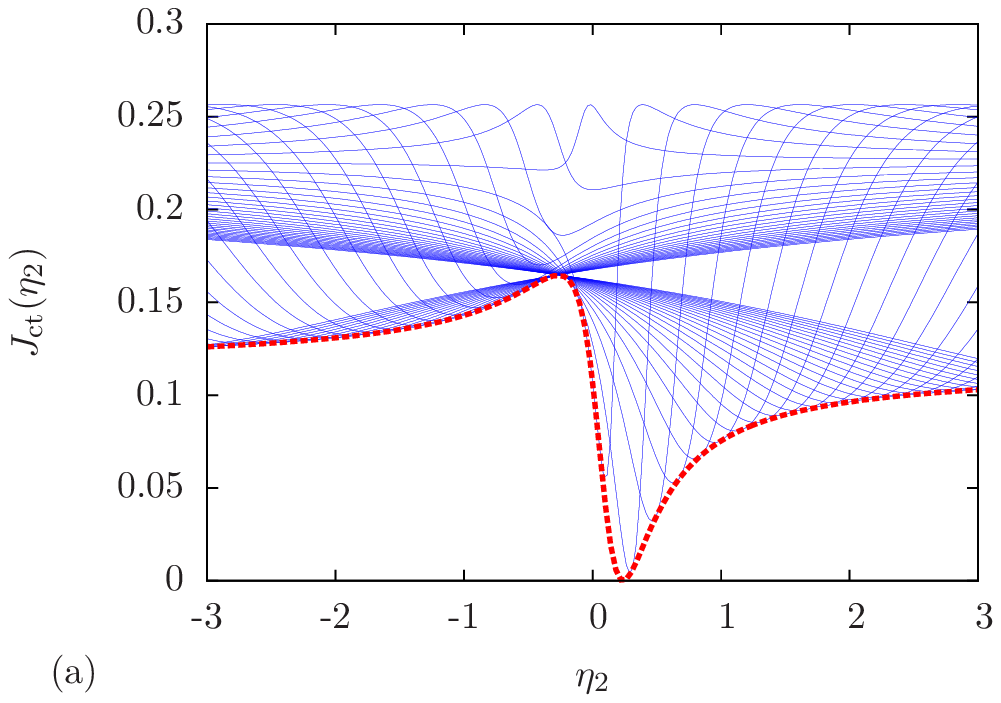}  \includegraphics[width=8cm]{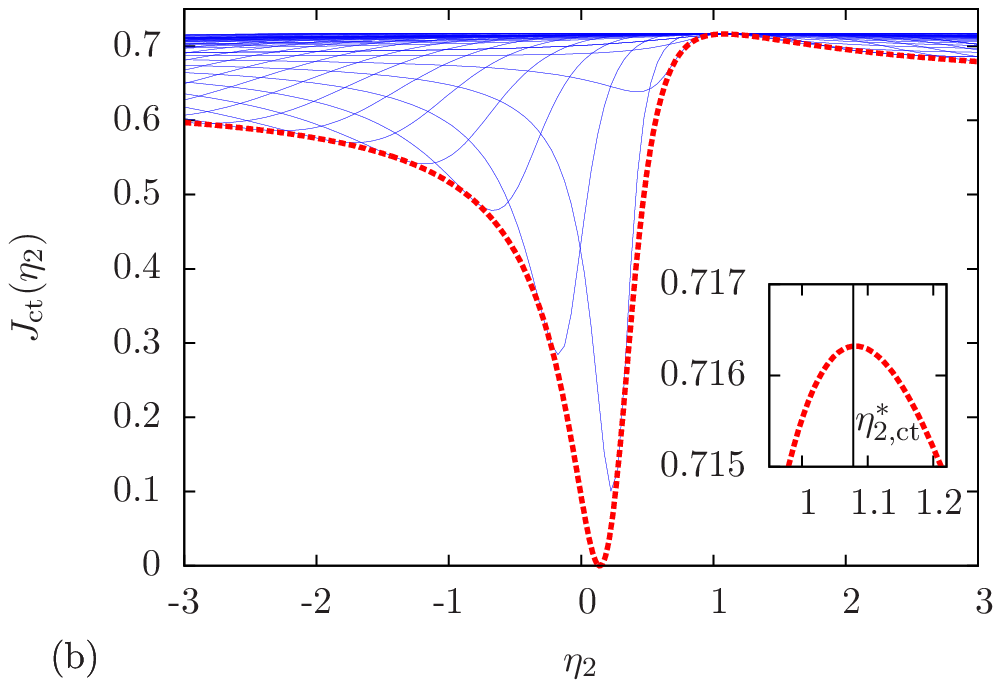} 
\caption{(a) Far-from-equilibrium contracted LDF $J_{\ct}(\eta_2)$ (thick dashed red line) and various crosssections of $J(\eta_2,\eta_3)$ (thin blue lines) for $\eta_3 \in [-10;10]$. The minimum is for $\bar{\eta}_2=0.24$. (b) LDF $J_{\ct}(\eta_2)$ contracted on small losses (thick dashed red line) and various crosssections of $J(\eta_2,\eta_3)$ (thin blue lines) for $\eta_3 \in [-10;10]$. The minimum is for $\bar{\eta}_2=0.14$ and the maximum for $\eta_{2,\ct}^* = 1.08$. The insert: zoom on the maximum.}
\label{FigContractionFE}
\end{figure*}

Finally, we comment on the effect of the contraction in Eq.~(\ref{minJc}) on the statistics of the remaining efficiency. This situation corresponds ignoring the third process even though it is still influencing the machine dynamics. In Fig.~\ref{FigContractionFE} we provide the contracted LDF $J_{\ct}(\eta_2)$. It displays the generic shape of an efficiency LDF except that no constraint exists on the position of the maximum, e.g. it is  below $\bar{\eta}_2$ in the FE case. This would be forbidden by the laws of thermodynamics in a machine with only two processes, but it is allowed whenever an additional process has been ignored in the description of the machine. Logically, when the ignored process is weakly irreversible as in the SL case of Fig.~\ref{FigContractionFE}(b), the maximum of the efficiency LDF must be located close to the reversible efficiency: In the limit of a vanishing affinity for the ignored process, we retrieve the usual efficiency fluctuations of a stationary machine with only two processes for which the reversible efficiency is the least likely.

\section{Conclusion}

In this paper, we focused on complex machines displaying not one but several goals. For each goal, we introduced an efficiency paying attention to thermodynamic consistency. Our motivation was twofold: first, stochastic machines with several goals may exist in nature. Second, if two different goals exist, one may affect the efficiency of the other one. We interpreted this as a loss in a machine with a single goal and analyzed the consequences of unknown losses on the efficiency statistics.

In these two cases, we described the general properties of the large deviation function of the efficiency, using the fluctuation theorem and assuming the convexity of the large deviation function for the entropy production, and we provided a method to obtain the large deviation function of the efficiency from the cumulant generating function of the entropy production. This paper extends the recent results of Refs.~\cite{Verley2014_vol5, Verley2014_vol90, Gingrich2014_vol16} on stochastic efficiency to the case of machines with more than two processes. In this case, we confirmed that the minimum of the large deviation function of the efficiency is still given by the macroscopic efficiencies defined as the ratio of mean entropy productions, and the maximum is still connected to an entropy production minimum. However, in the case of a machine with unknown losses, the least likely efficiency is reached for the most likely trajectories conditioned on the reversibility of the input and output processes. 

In the close-to-equilibrium limit we characterized the large deviation function of the efficiency using the response coefficient of the machine only (or equivalently using the entropy productions correlation functions). In this limit, we derived exactly the contracted large deviation function of the efficiency and found similarities with the efficiency fluctuations of a machine with only two processes. We support all our results by considering a simplified model of photoelectric cell. 

The theory developed in this paper includes the case of machines with an arbitrary number of processes: We provide in Appendices \ref{AppCGF} and \ref{AppLLE} the most important formula in the general case. Alternatively, one may always merge the various processes into two (or three) groups, the input processes and the output processes (and the loss processes) in order to use the theory developed for machines with two (or three) processes. This procedure is particularly convenient when considering that real physical systems often involve more than two processes, see, for instance, Ref.~\cite{Cuetara2015_vol17} about an electronic circuit composed of a double quantum dot channel capacitively coupled to a quantum point contact. In this reference the measurement of nanocurrents leads to non trivial interactions and additional dissipation in the device. At this point, beyond the number of processes required to model a machine, it is worth stressing that quantum coherence and destructive interference may significantly affect the fluctuations of the stochastic efficiency \cite{Esposito2015_vol91}.

\section*{Acknowledgment} 
 
We acknowledge H.-J. Hilhorst for his pertinent comments on this paper.

\appendix
\section{CUMULANT GENERATING FUNCTION}\label{AppCGF}
In this appendix, we obtain the efficiency LDF from the cumulant generating function of the entropy productions in the case of an arbitrary number $N$ of processes. We emphasize that this method can also be used to obtain the contracted LDF. We also remark that it is usually easier to compute numerically the efficiency LDFs using this method. 

The CGF $\phi $ and LDF $I$ for entropy productions are related by a Legendre transform
\begin{multline}
I(\sigma_{1},\sigma_{2},\ldots,\sigma_{N})\\
=\max_{\gamma_{1},\gamma_{2},\ldots,\gamma_{N}}\Big[\sum_{i=1}^{N}\gamma_{i}\sigma_{i}-\phi(\gamma_{1},\gamma_{2},\ldots,\gamma_{N})\Big].
\end{multline}
Introducing the efficiencies $\eta_i =-\sigma_i / \sigma_1$ for $i=2,\ldots ,N$, we can write
\begin{multline}
I(\sigma_{1},-\eta_2\sigma_{1},-\eta_3\sigma_{1},\ldots,-\eta_N\sigma_{1}) \\
=\max_{\gamma_{1},\gamma_{2},\ldots,\gamma_{N}}\Big[(\gamma_{1}-\sum_{i=2}^N\gamma_{i}\eta_i)\sigma_{1}-\phi(\gamma_{1},\gamma_{2},\ldots\gamma_{N})\Big]
\end{multline}
and the minimization of Eq. (\ref{MinJ}) gives
\begin{multline}
J(\eta_2,\eta_3,\ldots,\eta_N)\\
=\min_{\sigma_{1}}\max_{\gamma_{1},\gamma_{2},\ldots,\gamma_{N}}\Big[(\gamma_{1}-\sum_{i=2}^N\gamma_{i}\eta_i)\sigma_{1}-\phi(\gamma_{1},\gamma_{2},\ldots,\gamma_{N})\Big].
\end{multline}
We set $ \gamma = \gamma_{1}-\sum_{i=2}^N\gamma_{i}\eta_i $ to obtain
\begin{multline}
J(\eta_2,\eta_3,\ldots,\eta_N) =\min_{\sigma_{1}}\max_{\gamma}\bigg\{\gamma\sigma_{1} \\
+\max_{\gamma_{2},\ldots,\gamma_{N}}\Big[-\phi(\gamma+\sum_{i=2}^N\gamma_{i}\eta_i,\gamma_{2},\ldots,\gamma_N)\Big]\bigg\}.
\end{multline}
We now define the function,
\begin{eqnarray} \label{fetaalph}
f_{\eta_2,\ldots,\eta_N}(\gamma) &=&-\max_{\gamma_{2},\ldots,\gamma_N}\Big\{-\phi(\gamma+\sum_{i=2}^N\gamma_{i}\eta_i,\gamma_{2},\ldots,\gamma_N)\Big\} \nonumber \\
&=&\min_{\gamma_{2},\ldots,\gamma_N} \phi(\gamma+\sum_{i=2}^N\gamma_{i}\eta_i,\gamma_{2},\ldots,\gamma_N)
\end{eqnarray} 
and its Legendre transform,
\begin{equation}
\mathcal{F}_{\eta_2,\ldots,\eta_N}(\sigma_{1})= \max_{\gamma}\Big\{\gamma\sigma_{1}-f_{\eta_2,\ldots,\eta_N}(\gamma)\Big\}.
\end{equation}
Then the efficiency LDF can be rewritten
\begin{align}
J(\eta_2,\eta_3,\ldots,\eta_N)&=\min_{\sigma_{1}}\max_{\gamma}\Big\{\gamma \sigma_{1}-f_{\eta_2,\ldots,\eta_N}(\gamma)\Big\} \nonumber \\
&=\min_{\sigma_{1}}\mathcal{F}_{\eta_2,\ldots,\eta_N}(\sigma_{1}) \nonumber \\
&=-\max_{\sigma_{1}}\Big\{-\mathcal{F}_{\eta_2,\ldots,\eta_N}(\sigma_{1}) \Big\} \nonumber \\
&= -f_{\eta_2,\ldots,\eta_N}(0).
\end{align}
Using Eq.~(\ref{fetaalph}), we conclude that
\begin{equation}
J(\eta_2,\eta_3,\ldots,\eta_N)=-\min_{\gamma_{2},\ldots,\gamma_{N}} \phi\left(\sum_{i=2}^N\gamma_{i}\eta_i,\gamma_{2},\ldots,\gamma_{N}\right).
\end{equation}

\section{LEAST LIKELY EFFICIENCY}\label{AppLLE}

In this appendix we use the fluctuation theorem to prove some properties of the efficiency LDF in the general case of a machine with an arbitrary driving cycle and with $N$ processes contributing to the total entropy production. 

Along a contour line of the entropy productions' LDF, the total differential of $I$ vanishes,
\begin{align}
\dd I &= \sum_{i=1}^N \Dp{I}{\sigma_i}\dd \sigma_i \\
&=\dd \sigma_1 \left(\Dp{I}{\sigma_1} +\sum_{i=2}^N \Dp{I}{\sigma_i}\f{\dd \sigma_i}{\dd \sigma_1}\right) \\
	&= 0.
\end{align}
At the origin, we have $\eta_i^*=-\dd \sigma_i/\dd \sigma_1$ with $i=2,\ldots,N$ where the $\eta_i^*$'s are defined by $J(\eta_2^*,\ldots,\eta_N^*)= I(0,\ldots,0)$. So,
\begin{equation}
\sum_{i=2}^N \Dp{I}{\sigma_i}\bigg|_0\left(\Dp{I}{\sigma_1}\bigg|_0\right)^{-1} \eta_i^* = 1. \label{ConstOnLLEFF}
\end{equation}
We may repeat the arguments for the machine with the time-reversed driving cycle. We denote $\hat{I}(\sigma_1,\ldots,\sigma_N)$ as the entropy productions' LDF of this new machine and the efficiency LDF $\hat{J}(\eta_2,\ldots,\eta_N)$. If we define $\hat{\eta_i}^*$ by $\hat{J}(\hat{\eta}_2^*,\ldots,\hat{\eta}_N^*)= \hat{I}(0,\ldots,0)$, we have as above,
\begin{equation}
\sum_{i=2}^N \Dp{\hat{I}}{\sigma_i}\bigg|_0\left(\Dp{\hat{I}}{\sigma_1}\bigg|_0\right)^{-1} \hat{\eta}_i^* = 1.
\end{equation}
We now use the fluctuation theorem for the entropy productions:
\begin{equation}\label{FT}
I(\sigma_1,\ldots,\sigma_N) - \hat{I}(-\sigma_1,\ldots,-\sigma_N) = -\sum_{i=1}^N\sigma_i.
\end{equation}
Taking the partial derivatives of this equation at the origin yields
\begin{equation}\label{derivativeFT}
\Dp{I}{\sigma_i} \bigg|_0+\Dp{\hat{I}}{\sigma_i}\bigg|_0 = -1 \quad \text{with} \quad i \in \{1,\ldots,N\}.
\end{equation}
So, the least likely efficiencies of the machine with the time-reversed driving cycle are connected to those of the original machine. More specifically, for stationary machines or machines operating under time-symmetric driving for which $I(\sigma_1,\ldots,\sigma_N) = \hat{I}(\sigma_1,\ldots,\sigma_N)$, the least likely efficiencies satisfy the same constraint as the reversible efficiencies,
\begin{equation}
\sum_{i=2}^N \eta_i^* = 1,
\end{equation}
following from Eqs.~(\ref{ConstOnLLEFF}) and (\ref{derivativeFT}).

Furthermore evaluating the fluctuation theorem (\ref{FT}) at null entropy production, we have
\begin{multline}
I(\sigma_1,-\bar{\eta}_{2\text{ rev}}\sigma_1,\ldots,-\bar{\eta}_{N\text{ rev}}\sigma_1) \\ = \hat{I}(-\sigma_1,\bar{\eta}_{2\text{ rev}}\sigma_1,\ldots,\bar{\eta}_{N\text{ rev}}\sigma_1).
\end{multline}
which after minimization over $\sigma_1$ implies that the forward and reversed efficiency LDFs have the same values at reversible efficiencies,
\begin{equation}
J(\bar{\eta}_{2\text{ rev}},\ldots,\bar{\eta}_{N\text{ rev}}) = \widehat{J}(\bar{\eta}_{2\text{ rev}},\ldots,\bar{\eta}_{N\text{ rev}}). \label{SameValueAtRevEff}
\end{equation}
And still from (\ref{FT}) evaluated at the origin, we have $I(0,\ldots,0)= \widehat{I}(0,\ldots,0) $ , so the maximum of the forward and reversed efficiency LDFs have the same value, 
\begin{equation}
J(\eta_2^*,\ldots,\eta_N^*)=\widehat{J}(\widehat{\eta}_2^*,\ldots,\widehat{\eta}_N^*). \label{SameMaxValue}
\end{equation}
Let us emphasize that Eqs.~(\ref{SameValueAtRevEff}) and (\ref{SameMaxValue}) merge into the same equation for stationary machines or machines operating under time-symmetric driving since the least likely efficiencies become the reversible efficiencies.

\bibliography{Lien_ma_base_de_papier}

\end{document}